\documentclass[preprint,12pt]{elsarticle}

\usepackage{amsmath,amssymb,amsfonts}
\usepackage{graphicx}
\usepackage{booktabs}
\usepackage{lineno}
\usepackage{array}

\journal{Nuclear Physics B}

\begin{document}

\begin{frontmatter}

\title{From Gradient Postulates to Gauge Structure:\\
Chirality, Conserved Charges, and Constraints on Internal Symmetry}

\author[lhep]{Xinqiao Li\corref{cor1}}
\ead{lixq@ihep.ac.cn}

\affiliation[lhep]{State Key Laboratory of Particle Astrophysics, Institute of High Energy Physics, Chinese Academy of Sciences, Beijing 100049, China}

\cortext[cor1]{Corresponding author}

\begin{abstract}
We propose a conceptual framework in which, in the non-relativistic static limit, conservative forces arise as gradients of guidance potentials, and the charges that generate these potentials are conserved and form a Lie algebra. We examine what constraints these postulates, together with the empirical facts of parity violation and non-Abelian weak isospin, impose on the internal symmetry group of a gauge theory. We argue that the non-Abelian character of the gauge sector follows from the conserved-charge algebra together with the observed non-Abelian weak isospin algebra. The requirement that the internal group accommodate chiral fermions constrains the admissible representations. Classical stability, charge quantization, and anomaly cancellation provide three physical consistency conditions that strongly constrain the internal group to be compact. With a minimality criterion, the Standard Model group emerges as the minimal candidate among a small set of compact anomaly-free groups with chiral content. We explicitly identify the minimality criterion as a selective assumption rather than a derived result. We do not claim a unique derivation of the Standard Model group. We also show that the scalar guidance potential reproduces Newtonian gravity and Coulomb's law in the non-relativistic limit, and that relativistic consistency requires its upgrade to a metric tensor and gauge vector fields. The postulates substantially constrain the space of viable gauge theories and provide a conceptual basis for treating the gauge group as a constrained rather than arbitrary input.
\end{abstract}


\end{frontmatter}

\section{Introduction}
\label{sec:intro}

Newtonian mechanics treated forces as action at a distance between bodies~\cite{newton1687}. The revolutions of Maxwell and Einstein recast interactions as local transmissions through fields or spacetime geometry~\cite{maxwell1865,einstein1916}. Gravitation, electromagnetism, the strong force, and the weak force still rely on distinct mathematical structures, lacking a unified intuitive principle.

The origin of the Standard Model gauge group $SU(3)_c \times SU(2)_L \times U(1)_Y$ remains an open conceptual question. Grand unified theories embed it in a larger group~\cite{georgi1974}, but the choice of the embedding group is itself not explained by the postulates of the Standard Model. Approaches based on swampland constraints argue for certain gauge group structures~\cite{ooguri2007}, but they rely on string-theoretic consistency conditions rather than dynamical first principles.

This paper asks whether a small set of postulates about the nature of force and charge can constrain the gauge group rather than assume it. We try to show that the postulates P1--P3 substantially constrain the space of viable gauge theories, and that the Standard Model group appears as the minimal candidate under a selective minimality criterion.

We propose three postulates. The first is:
\begin{enumerate}
\item[(P1)] In the non-relativistic static limit, conservative elementary forces arise as gradients of guidance potentials. Relativistic covariance requires the guidance potentials to be upgraded to a metric tensor and gauge vector fields, which may be assembled into a total connection $\mathcal{A}=\Gamma+A$. The curvature $\mathcal{F}=d\mathcal{A}+\mathcal{A}\wedge\mathcal{A}$ organizes the gravitational and gauge sectors. In the minimal realization considered here, the two sectors take values in commuting factors, and a more general connection with non-trivial mixing is left for future work.
\end{enumerate}

Physically, the guidance potentials represent the local deformation of space. A charge deforms the space around it, and the guidance potential quantifies this deformation. Its gradient drives the motion of other charges. This picture provides a local static description of forces. For gravity, the deformation is that of the external spacetime metric. For gauge interactions, the deformation is that of an internal space; the precise geometric realization is left for future work. In the relativistic completion, the total connection organizes the spacetime connection $\Gamma$ and the internal gauge connection $A$ within a single geometric object.

The second postulate captures the structure of conserved charges:
\begin{enumerate}
\item[(P2)] Every conserved internal charge $Q^a$ generates a local guidance field $A_\mu^a$, with the conserved current $J^{a\mu}$ as its source. The charges form a finite-dimensional Lie algebra
\begin{equation}
[Q^a,Q^b]=i f^{abc}Q^c,
\end{equation}
and at least one non-Abelian subalgebra exists. Experimentally, the weak isospin charges $T^i$ satisfy
\begin{equation}
[T^i,T^j]=i\epsilon^{ijk}T^k,
\end{equation}
so the weak charge algebra is non-Abelian. Localization requires
\begin{equation}
\partial_\mu \to D_\mu = \partial_\mu + ig A_\mu^a T^a,
\end{equation}
and the field strength acquires the non-linear term
\begin{equation}
F_{\mu\nu}^a = \partial_\mu A_\nu^a - \partial_\nu A_\mu^a + g f^{abc} A_\mu^b A_\nu^c .
\end{equation}
Physical charge spectra must be discrete.
\end{enumerate}

P1 and P2 together are sufficient to recover Newtonian gravity and Coulomb electrostatics in the non-relativistic limit. They also show that relativistic consistency requires the guidance potentials to be upgraded to a spacetime metric and gauge vector fields. Furthermore, the non-Abelian part of the charge algebra directly generates non-Abelian gauge fields.

The third postulate captures a fundamental empirical fact:
\begin{enumerate}
\item[(P3)] The internal symmetry group $G_{\rm int}$ must admit chiral fermions, i.e., left- and right-handed components transform in inequivalent complex representations. The chiral representation must satisfy four-dimensional perturbative anomaly cancellation. The experimental basis is parity violation in the weak interaction, the left-handedness of neutrinos, and the $V-A$ structure of the weak force.
\end{enumerate}

P3 is motivated by the parity violation experiments of the 1950s~\cite{lee1956,wucs1957}. It elevates the observed chirality of weak interactions to a foundational postulate. This elevation follows the historical precedent of the equivalence principle, which was likewise an empirical regularity elevated to a foundational postulate.

P3 is a postulate, not a theorem. Its justification is empirical. Its value is that it constrains the space of possible theories.

Elevating chirality to the status of a postulate is a key step in the logical architecture of the theory. In the development of physical theories, it is common for empirically observed regularities to be promoted to fundamental principles. The constancy of the speed of light, the equivalence of inertial and gravitational mass, and the quantization of electric charge all entered their respective theories in this manner. Postulate 3 follows this tradition. By requiring that the internal symmetry group accommodate chiral fermions and satisfy anomaly cancellation, it constrains the admissible representations and, together with P2, strongly restricts the internal group. The parity violation discovered in the 1950s is thus not an incidental property of weak interactions, but a manifestation of the intrinsic chirality of the internal space. The non-Abelian nature of the gauge sector, the chiral fermion content of the Standard Model, and the observed violation of parity are traced to the combination of P2 and P3.

The logical structure of the paper is as follows:
\begin{enumerate}
\item P1 gives the non-relativistic guidance potential and its relativistic upgrade to a metric tensor and gauge fields.
\item P2 gives the conserved-charge algebra and, via the observed non-Abelian weak isospin algebra, the non-Abelian gauge structure.
\item P3 gives chirality and anomaly cancellation, which constrain the admissible representations.
\item Classical stability, charge quantization, and anomaly cancellation strongly constrain the internal group to be compact.
\item Compactness, anomaly cancellation, and a minimality criterion narrow the candidates to a small set, with the Standard Model group as the minimal candidate.
\end{enumerate}

The paper develops a non-relativistic scalar guidance field and identifies its limitations. We show that relativistic completion requires the scalar potential to be upgraded to a metric tensor and gauge vector fields. From the three postulates, the framework yields a non-Abelian gauge structure. Physical consistency conditions strongly constrain the internal symmetry group to be compact. Combined with anomaly cancellation and a minimality criterion, this narrows the admissible gauge groups to a small set, with the Standard Model group as the minimal candidate.

\section{Non-relativistic scalar guidance field}
\label{sec:scalar}

\subsection{Basic equations and the superposition principle}

For every interaction we introduce a physical quantity $P$, called a ``charge''. A point charge at rest generates a static scalar guidance potential $\Phi(\mathbf{r})$. A test body carrying charge $p$ experiences the force
\begin{equation}
\mathbf{F} = -p\,\nabla\Phi .
\label{eq:force}
\end{equation}
The negative sign ensures that a positive test charge moves toward lower potential. The potential satisfies a Poisson equation sourced by the charge density $\rho_P$:
\begin{equation}
\nabla^2 \Phi = 4\pi\alpha\,\rho_P ,
\label{eq:poisson}
\end{equation}
where $\alpha$ is a coupling constant characteristic of the interaction. For a point charge of magnitude $P$ at the origin, the solution is
\begin{equation}
\Phi(r) = -\frac{\alpha P}{r}.
\label{eq:point_potential}
\end{equation}
The flux of the gradient through any closed surface is independent of the radius, leading to an inverse-square law in the static limit.

Potentials generated by different charges obey a linear superposition principle:
\begin{equation}
\Phi(\mathbf{r}) = \sum_i \frac{-\alpha P_i}{|\mathbf{r}-\mathbf{r}_i|}.
\label{eq:superpos}
\end{equation}
This expresses mathematically the absence of direct interaction among the charges in the static limit.

\subsection{Recovery of gravity and electrostatics}

To recover Newtonian gravity we choose $\alpha=G$ and $P=M$ (gravitational mass), where $G$ is the Newtonian gravitational constant. The potential is
\begin{equation}
\Phi_g(r) = -\frac{GM}{r},
\end{equation}
and the force on a test mass $m$ is
\begin{equation}
\mathbf{F}_g = -m\nabla\Phi_g = -\frac{GMm}{r^2}\hat{r},
\end{equation}
which is attractive for positive masses.

To recover electrostatics we choose a separate symbol $\phi_e$ for the electric potential, with
\begin{equation}
\phi_e(r) = \frac{Q}{4\pi\varepsilon_0 r}.
\end{equation}
The force on a test charge $q$ is
\begin{equation}
\mathbf{F}_e = -q\nabla\phi_e = \frac{qQ}{4\pi\varepsilon_0 r^2}\hat{r},
\end{equation}
which is repulsive for like charges and attractive for opposite charges. Attraction and repulsion are unified through the sign of the charge, not through the sign of the scalar potential alone. The gravitational and electrostatic conventions differ in the sign of the source term; they should not be identified without qualification. In the gravitational case, the source is the mass $M$ and the coupling $G$ is positive, so the potential is negative and the force is attractive. In the electrostatic case, the source is the charge $Q$, which can be positive or negative, so the potential and force can have either sign.

\subsection{Limitations of the scalar description}

The scalar guidance field possesses three fundamental limitations:
\begin{enumerate}
\item No magnetism. The gradient of a scalar cannot produce a velocity-dependent Lorentz force.
\item Incorrect light deflection. A pure scalar gravity theory gives the wrong bending of starlight~\cite{nordstrom1913}.
\item Source mismatch. In relativity the gravitational source is the second-rank energy-momentum tensor $T^{\mu\nu}$, with which a scalar cannot couple covariantly.
\end{enumerate}
Each of these limitations directly motivates the relativistic upgrades of Sec.~\ref{sec:relativistic}.

\section{Relativistic completion}
\label{sec:relativistic}

\subsection{Gravity: why a metric is natural}

A theory of gravity must be compatible with special relativity and act universally on all forms of energy. A standard and natural way to satisfy both requirements is to promote the scalar potential $\Phi$ to the time-time component of a metric tensor $g_{\mu\nu}$. In the weak-field limit, using the metric signature $(-,+,+,+)$,
\begin{equation}
g_{00} \approx -(1+2\Phi/c^2),\qquad g_{ij} \approx (1-2\Phi/c^2)\,\delta_{ij}.
\end{equation}
The gravitational source $T^{\mu\nu}$ is a symmetric second-rank tensor. The guidance field must be a second-rank tensor to couple consistently. This leads to the Einstein-Hilbert action and to the Einstein field equations
\begin{equation}
G_{\mu\nu} = \frac{8\pi G}{c^4} T_{\mu\nu},
\end{equation}
where $G_{\mu\nu}$ is the Einstein tensor.

\subsection{Electromagnetism: why a vector field is natural}

Electromagnetism includes velocity-dependent magnetic forces. To produce a four-dimensional Lorentz force
\begin{equation}
f^\mu = q\,F^{\mu\nu}_{\rm em}\,u_\nu ,
\end{equation}
an antisymmetric tensor $F_{\mu\nu}^{\rm em}$ is required, where $q$ is the electric charge and $u_\nu$ is the four-velocity. The minimal construction introduces a four-vector potential $A_\mu^{\rm em}$ and defines
\begin{equation}
F_{\mu\nu}^{\rm em} = \partial_\mu A_\nu^{\rm em} - \partial_\nu A_\mu^{\rm em} .
\end{equation}
Gauge invariance is implemented via minimal coupling $\partial_\mu \to \partial_\mu + (iq/\hbar) A_\mu^{\rm em}$. Thus the upgrade from scalar to vector is strongly suggested by relativistic covariance and the existence of magnetic forces.

\section{From postulates to non-Abelian gauge structure}
\label{sec:nonabelian}

\subsection{Non-Abelian structure from conserved charges}

Postulate P2 requires that every conserved internal charge $Q^a$ generates a local guidance field $A_\mu^a$. The charges form a finite-dimensional Lie algebra,
\begin{equation}
[Q^a, Q^b] = i f^{abc} Q^c .
\end{equation}
Experimentally, the weak isospin charges $T^i$ satisfy
\begin{equation}
[T^i, T^j] = i\epsilon^{ijk} T^k ,
\end{equation}
so the weak charge algebra is non-Abelian. This is an empirical fact, independent of any representation-theoretic argument about chirality.

When the global charge algebra is localized, the covariant derivative becomes
\begin{equation}
D_\mu = \partial_\mu + i g A_\mu^a T^a ,
\end{equation}
and the field strength acquires the non-linear term~\cite{yang1954}
\begin{equation}
F_{\mu\nu}^a = \partial_\mu A_\nu^a - \partial_\nu A_\mu^a + g f^{abc} A_\mu^b A_\nu^c .
\label{eq:nonab}
\end{equation}
The $A_\mu^a$ constitute a connection on a principal fiber bundle with structure group $G_{\rm int}$ generated by the $Q^a$.

Thus the non-Abelian character of the gauge sector follows from P2, together with the experimental fact that the weak isospin algebra is non-Abelian. It does not follow from chirality alone.

\subsection{Chirality and representation theory}

Postulate P3 requires that the internal symmetry group $G_{\rm int}$ admit chiral fermions: left- and right-handed components transform in inequivalent complex representations. An Abelian group such as $U(1)$ admits complex one-dimensional representations $\rho_n(e^{i\theta})=e^{in\theta}$ with $n\neq 0$, and these are not equivalent to their complex conjugates. Therefore chirality alone does not imply non-Abelianity. What P3 does imply is that the admissible representations must be complex and that the fermion content must satisfy anomaly cancellation.

For a non-Abelian group, complex representations are generic, although not universal. $SU(2)$, for instance, has only real or pseudoreal irreducible representations. The Standard Model nevertheless realizes chirality through reducible $SU(2)_L$ representations: left-handed fermions transform as doublets, right-handed fermions as singlets. Chirality is therefore a property of the full reducible representation content, not of any single irreducible representation. For an Abelian group, chirality can be realized by assigning different $U(1)$ charges to left- and right-handed fermions. Thus P3 constrains the representation content but does not by itself force the group to be non-Abelian. The non-Abelian structure comes from P2.

\subsection{The total connection}
\label{sec:totalconn}

The gauge potentials $A_\mu^a$ introduced above can be viewed as components of a more general geometric object. Let $\Gamma$ be the spacetime connection valued in the Lorentz Lie algebra $\mathfrak{so}(3,1)$, and let
\begin{equation}
A = A_\mu^a T^a dx^\mu
\end{equation}
be the internal gauge connection valued in a candidate internal Lie algebra $\mathfrak{g}_{\rm int}$ to be constrained below. The generators $T^a$ satisfy the algebra $[T^a, T^b] = i f^{abc} T^c$.

We combine the two connections into a total connection
\begin{equation}
\mathcal{A} = \Gamma + A ,
\label{eq:totalconn}
\end{equation}
valued in the direct sum $\mathfrak{so}(3,1)\oplus\mathfrak{g}_{\rm int}$. The total curvature is
\begin{equation}
\mathcal{F} = d\mathcal{A} + \mathcal{A} \wedge \mathcal{A} .
\label{eq:totalcurv}
\end{equation}
In the minimal realization considered in this paper, $\Gamma$ and $A$ take values in commuting factors of a direct sum, so that
\begin{equation}
\Gamma\wedge A + A\wedge\Gamma = 0 .
\end{equation}
Hence
\begin{equation}
\mathcal{F} = \mathcal{R} + F ,
\label{eq:decomp}
\end{equation}
where $\mathcal{R}=d\Gamma+\Gamma\wedge\Gamma$ is the Riemann curvature two-form and $F = dA + A \wedge A$ is the gauge field strength two-form.

In this minimal realization, the two sectors remain geometrically distinct: gravity is described by the $\mathfrak{so}(3,1)$-valued component $\Gamma$, while gauge interactions are described by the $\mathfrak{g}_{\rm int}$-valued component $A$. A more general total connection with non-trivial mixing between the two sectors, in which the cross terms $\Gamma\wedge A + A\wedge\Gamma$ no longer vanish, is left for future work. The total-connection language is therefore used here as a structural device: it organizes the gravitational and gauge sectors within a common geometric framework, without claiming that they are dynamically unified in the minimal realization. The question of whether a non-trivial mixing term can arise from a more fundamental theory is left open.

\section{Constraints toward internal compactness}
\label{sec:compactness}

We now collect three physical consistency conditions that strongly constrain the internal symmetry group $G_{\rm int}$ to be compact. These are physical arguments, not mathematical theorems.

\subsection{Classical stability}

The Yang-Mills Lagrangian yields kinetic terms involving the Cartan-Killing metric
\begin{equation}
\kappa^{ab} = f^{acd} f^{bcd} .
\end{equation}
For a semisimple Lie group, the energy is bounded below only if $\kappa^{ab}$ is definite. A classic theorem states that a semisimple Lie group admits a definite Killing metric if and only if it is compact~\cite{weinberg1996,knapp2002}. Non-compact semisimple groups have indefinite metrics. They lead to ghost modes and classical instability.

For an Abelian factor such as $U(1)$, the Killing form vanishes identically. The compactness of the $U(1)$ factor is therefore not established by this argument. It follows instead from the discreteness of the charge spectrum, as discussed in Sec.~\ref{sec:discrete}.

\subsection{Discrete representation theory and charge quantization}
\label{sec:discrete}

A second condition comes from the representation theory of the internal group. The observed quantization of electric charge requires that the charges carried by physical states take discrete values. This is not automatic. It requires the internal symmetry group to possess a discrete spectrum of unitary representations.

Compact Lie groups have exactly this property. Their unitary representations are finite-dimensional and labeled by discrete integers. For instance, the compact $U(1)$ group has representations $e^{in\theta}$ with integer $n$, giving charge quanta $ne$. The representations of compact non-Abelian groups are similarly labeled by discrete highest weights.

Non-compact groups typically lack the finite-dimensional unitary representations needed to construct a finite fermion spectrum. Their unitary representations are either infinite-dimensional, or trivial on the non-compact directions. Constructing a consistent quantum gauge theory with a finite set of charge quanta carried by finite-dimensional representations becomes difficult.

This representation-theoretic argument is independent of any quantum tunneling process. It concerns the algebraic structure of the group, not its dynamics. It therefore remains valid within the classical framework of the present paper.

\subsection{Anomaly cancellation}
\label{sec:anomaly}

A third condition comes from the cancellation of gauge anomalies. In four dimensions, the anomaly coefficient for a chiral gauge theory involves group-theoretic invariants of the gauge group~\cite{adler1969,bell1969,bouchiat1972,gross1972}. Anomaly cancellation is a quantum consistency condition. It requires the fermion content of the theory to satisfy precise algebraic identities.

Anomaly cancellation is an algebraic condition that can be formulated for any Lie algebra. However, the discrete representation theory of compact groups provides the natural framework in which these conditions take a precise and manageable form. The relevant anomaly coefficients are sums over the discrete weights of the fermion representations. Non-compact groups typically lack the required finite-dimensional representation-theoretic framework, making it difficult to accommodate the precisely balanced fermion assignments observed in nature.

The Standard Model fermion content satisfies all anomaly cancellation conditions exactly. This is a non-trivial consistency check. It is not known how to construct a consistent non-compact chiral gauge theory with the observed discrete charge spectrum and anomaly cancellation.

We emphasize that this argument does not invoke instantons or any other non-perturbative topological object. It concerns the algebraic structure of the fermion representation content.

\subsection{The internal space as a compact group manifold}
\label{sec:internal_geom}

The compactness argument acquires a natural geometric interpretation. The internal symmetry group $G_{\rm int}$ acts as a group of isometries on a compact internal space $K$. The gauge fields $A_\mu^a$ are the components of the total connection along the Killing vectors of $K$. In the simplest realization, $K$ is itself a group manifold and $G_{\rm int}$ acts on it by left translations; the full isometry group may be larger, but the relevant structure group for the gauge fields is $G_{\rm int}$.

The internal space $K$ carries a bi-invariant metric
\begin{equation}
h_{ab} = \frac{1}{\lambda^2} \kappa_{ab} ,
\label{eq:biinv}
\end{equation}
where $\kappa_{ab}$ is the Cartan-Killing metric on $\mathfrak{g}_{\rm int}$ and $\lambda$ is an inverse length scale. Compactness of $K$ guarantees finite volume and a finite Haar measure. While non-compact spaces can also have finite volume, they typically lack the discrete unitary representation theory that charge quantization and anomaly cancellation require. Compactness is therefore the natural setting that simultaneously satisfies all three consistency conditions.

The internal space carries topological defects. These defects are classified by the homotopy groups of $K$. For a compact group manifold, the relevant homotopy groups are discrete. For instance, $\pi_3(SU(3)) = \mathbb{Z}$ and $\pi_3(SU(2)) = \mathbb{Z}$. These discrete classifications provide a natural geometric setting for quantized charges, though the detailed identification with observed charges is not attempted here.

The geometric picture of the internal space also suggests the origin of the Standard Model gauge group. If the internal manifold factorizes as
\begin{equation}
K = SU(3) \times SU(2) \times U(1) ,
\label{eq:K}
\end{equation}
the three factors correspond to three independent sectors of the internal geometry. Their product structure is the geometric origin of the direct-product structure of the Standard Model gauge group.

\subsection{Summary of the compactness argument}

Combining the three conditions, we conclude that the internal symmetry group $G_{\rm int}$ is strongly constrained to be a compact Lie group. The three conditions are:

\begin{enumerate}
\item \textbf{Classical stability.} For the semisimple part, the energy is bounded below only if the Cartan-Killing metric is definite, which requires compactness. For the Abelian part, compactness follows from charge quantization.
\item \textbf{Charge quantization.} Discrete charges require discrete unitary representations, which compact groups provide and non-compact groups typically do not.
\item \textbf{Anomaly cancellation.} The precise formulation of anomaly cancellation conditions in chiral gauge theories relies on the discrete representation theory of compact groups.
\end{enumerate}

The three conditions are not logically independent: charge quantization and anomaly cancellation both rely on discrete representation theory. They are nevertheless physically distinct, since they arise from different consistency requirements.

None of these arguments depends on quantum tunneling, instantons, or magnetic monopoles. They are structural properties of compact Lie groups that hold within the classical geometric framework of this paper. We treat compactness as a well-motivated physical constraint, not as a mathematical theorem.

According to the classification theorem, connected compact Lie groups are, up to finite covers, products of $U(1)$ factors and compact simple Lie groups $SU(n)$, $SO(n)$, $Sp(n)$, and the five exceptional groups~\cite{weinberg1996,georgi1982}.

\section{Narrowing the admissible gauge groups}
\label{sec:sm}

The classification of compact Lie groups provides a finite list of candidates, but does not by itself select the Standard Model group. The additional input comes from the requirement of chiral fermions and anomaly cancellation (P3), together with the minimality criterion.

Compactness alone leaves many candidates. P3 requires chiral fermions, which in turn demands anomaly cancellation. In four dimensions, the anomaly coefficients involve group-theoretic invariants of the gauge group. One generation of quarks and leptons under $SU(3)_c \times SU(2)_L \times U(1)_Y$ exactly cancels all gauge anomalies~\cite{bouchiat1972,gross1972}.

Several compact anomaly-free groups with chiral content are known. We summarize a representative set in Table~\ref{tab:candidates}. The table is not exhaustive, but it illustrates the range of possibilities and the role of the minimality criterion.

\begin{table*}[t]
\centering
\caption{Representative compact anomaly-free groups with chiral content. The column ``Minimal chiral content'' lists the minimal anomaly-free chiral representation. The column ``Status'' summarizes the main reason for disfavoring or favoring the candidate under the minimality criterion.}
\label{tab:candidates}
\begin{tabular}{@{}llll@{}}
\toprule
Candidate group & Rank & Minimal chiral content & Status under minimality \\
\midrule
$SU(3)\times SU(2)\times U(1)$ & 4 & 15 Weyl fermions & Minimal \\
$SU(5)$ & 4 & $\bar{\mathbf{5}}\oplus\mathbf{10}$ & Additional high-dimensional \\
 & & & adjoint scalar content \\
$SO(10)$ & 5 & $\mathbf{16}$ spinor & Rank larger \\
$E_6$ & 6 & $\mathbf{27}$ & Rank larger \\
\bottomrule
\end{tabular}
\end{table*}

\subsection{The minimality principle}

To select among compact anomaly-free groups, we invoke a minimality principle: minimize the total rank of the gauge group and the number of chiral matter representations.

The rank of $SU(3)_c\times SU(2)_L\times U(1)_Y$ is 4. The minimal anomaly-free representation consists of one quark-lepton generation (15 Weyl fermions). In the examples considered here, grand unified groups typically increase either the rank or the representation complexity. For example, $SU(5)$ has rank 4 but requires additional high-dimensional scalar content for symmetry breaking. $SO(10)$ has rank 5 and spinorial matter. $E_6$ has rank 6 and a 27-dimensional representation.

Under the combined criteria of compactness, anomaly freedom, and minimality, $SU(3)_c\times SU(2)_L\times U(1)_Y$ emerges as the minimal candidate among the set of compact anomaly-free groups with chiral content.

We present this as a selective principle, not as a theorem derived from the postulates. The minimality criterion is a selective assumption. We do not claim that this choice is unique.

\subsection{Status of the selection}

The minimality criterion provides a natural organizing principle but does not by itself single out a unique group. Other compact anomaly-free groups with chiral content, such as $SO(10)$ or $E_6$, are not excluded by the postulates. They are disfavored only by the minimality criterion, which is a selective assumption rather than a derived result.

A first-principles derivation of the minimality criterion from the postulates remains an open problem. In the present framework, the Standard Model group is best regarded as the minimal natural candidate among a small set of compact anomaly-free groups, rather than as a uniquely derived result.

\section{Discussion}
\label{sec:discussion}

\subsection{What has been constrained}

The postulates P1, P2, and P3 substantially constrain the internal symmetry group. P2, together with the experimental fact of non-Abelian weak isospin, requires a non-Abelian gauge group. Classical stability, charge quantization, and anomaly cancellation strongly constrain the group to be a compact Lie group. Anomaly cancellation and a minimality criterion narrow the admissible gauge groups to a small set. The Standard Model group appears as the minimal candidate within this set.

The gauge group is therefore constrained by the postulates rather than freely chosen. This paper does not claim to derive the Standard Model group uniquely, nor does it construct a complete unified theory.

\subsection{What remains open}

We have not proven that the Standard Model group is uniquely derived from the postulates. The minimality principle is a selective assumption. The compactness argument is a physical constraint, not a mathematical theorem. These are open problems.

Three derivation paths can be considered. The first uses the representation theory of compact Lie groups. The second uses the geometry of the internal manifold. The third uses the framework of geometric quantization. A first-principles derivation of the minimality criterion remains the central open problem.

\subsection{Comparison with earlier approaches}

Kaluza-Klein theory derives gauge fields from extra dimensions~\cite{kaluza1921,klein1926}, but it does not constrain the gauge group. String theory determines gauge groups by world-sheet consistency on compact manifolds, but it relies on additional geometric structures. Approaches based on swampland constraints derive gauge group structures from string-theoretic consistency conditions~\cite{ooguri2007}.

The present framework is different. It derives the non-Abelian character of the gauge group from the conserved-charge algebra of P2, together with the empirical fact of non-Abelian weak isospin. It constrains compactness from physical consistency. It does not rely on string-theoretic consistency conditions. The present paper works within four-dimensional gauge theory, while using the total-connection language as a structural device. A fully higher-dimensional geometric realization is left for future work.

\subsection{Outlook}
\label{sec:outlook}

The conceptual framework developed in this paper provides a basis for a more complete geometric realization. The internal symmetry group $G_{\rm int}$ acts as a structure group on a compact internal manifold $K$. The gauge fields are components of a total connection on a product space. This geometric picture may provide a unifying perspective on gravity and gauge interactions.

The non-Abelian gauge structure derived here applies to both the strong and weak sectors. The strong interaction corresponds to the $SU(3)_c$ factor. The weak interaction corresponds to the $SU(2)_L$ factor. Their detailed dynamics, including confinement and the Higgs mechanism, require additional structure beyond the present framework.

A more complete geometric realization of the total-connection framework, including its dynamical consequences, is left for future work.

Several open problems remain. The minimality principle requires a first-principles derivation from the postulates. The compactness argument, while physically compelling, is not a mathematical theorem. A more complete geometric realization of the total connection is left for future work.

\section*{Acknowledgments}
The author thanks all those who have provided inspiration and assistance. This work was supported by grants from the National Natural Science Foundation of China (NNSFC) (grant No.~12673120,~12173038).

\end{document}